\documentclass[a4paper]{article}
\usepackage{anyfontsize}
\usepackage{INTERSPEECH2019}
\usepackage[table]{xcolor}
\usepackage{booktabs}
\usepackage{graphicx}
\usepackage[noadjust]{cite}
    
\usepackage{hyperref}
\usepackage{caption}

\usepackage{float}
\title{  SPEECH EMOTION RECOGNITION USING QUATERNION CONVOLUTIONAL NEURAL NETWORKS  }
\name{Aneesh Muppidi and Martin Radfar,}
\address{Department of Computer Science, Stony Brook University, YN, USA }
\email{aneeshmuppidi19@gmail.com, radfar@cs.stonybrook.edu}

\begin{document}

\maketitle
\begin{abstract}

Although speech recognition has become a widespread technology, inferring emotion from  speech signals still remains a challenge. To address this problem, this paper proposes a quaternion convolutional neural network (QCNN) based speech emotion recognition (SER) model in which Mel-spectrogram features of speech signals are encoded in an RGB quaternion domain. We show that our QCNN based SER model outperforms other real-valued methods in the Ryerson Audio-Visual Database of Emotional Speech and Song (RAVDESS, 8-classes) dataset, achieving, to the best of our knowledge, state-of-the-art results. The QCNN also achieves comparable results with the state-of-the-art methods in the Interactive Emotional Dyadic Motion Capture (IEMOCAP 4-classes) and Berlin EMO-DB (7-classes) datasets. Specifically, the model achieves an accuracy of 77.87\%, 70.46\%, and 88.78\% for the RAVDESS, IEMOCAP, and EMO-DB datasets, respectively. In addition, our results show that the quaternion unit structure is better able to encode internal dependencies to reduce its model size significantly compared to other methods. 
\end{abstract}
\noindent\textbf{Index Terms}: Speech Emotion Recognition, Signal Processing, Quaternion Deep Learning, Convolutional Neural Networks

\section{Introduction}
Speech emotion recognition (SER) is an active, yet challenging area of research that has many important implications in  technologies such as automated healthcare, clinical trials, voice assistants, psychological therapy, emergency responders, call centers, video games, robot-human interactions, and more. Nonetheless, recognizing emotion from speech signals is difficult due to many reasons, namely: qualitative properties of identifying emotion, background noise, variable individual-specific accentuation, weak representation of grammatical and semantic knowledge, temporal and spectral characteristics of the signal due to fast or slow speech, dynamical characteristics such as soft or loud voices, and multiple voices \cite{rethage2018wavenet, benzeghiba2007automatic, ramakrishnan2013speech}. In addition, if an application can learn high-level features from speech for emotion classification, it is possible to replicate or imitate emotion from a text-to-speech context, further improving the aforementioned fields of impact  \cite{ashar2020speaker}.
	
In general, SER is a classification task in which we extract features from a set of emotion-labeled speech signals and use the pair of feature and label to train a classifier which is commonly a deep neural architecture.   Features such as MFCC, chromograms, spectral contrast features, and Tonnetz representations have been used in previous neural based SER models  \cite{an2017emotional, pascual2017segan, humphrey2012learning}. Neural based SER models usually leverage n-dimensional convolutional neural networks (CNNs) \cite{palaz2015convolutional}, recurrent neural networks (RNNs) \cite{miao2015eesen}, Long short-term memory networks (LSTMs) \cite{graves2013hybrid}, or as combinations and variations of these techniques \cite{wollmer2010combining}. 

Recently, the quaternion based neural networks have been shown  to better encode features than the standard real-valued based neural networks \cite{isokawa2003quaternion}. The use of the quaternion domain has shown a unique ability in color image processing to capture all three channels of the RGB domain without the typical information loss of separating these channels, while reducing the size of its model in comparison to real-valued models \cite{zhu2018quaternion}.

In this paper, to the best of our knowledge, we propose the first quaternion based model that is able to encode an RGB representation of Mel-spectrograms for the application of SER classification. Using this approach, we propose a quaternion convolutional neural network (QCNN) to train on benchmark SER datasets in the quaternion domain with quaternion converted elements of a CNN. 
	The results, conducted on three public SER datasets, show that the QCNN model is able to yield state-of-the-art results on RAVDESS and outperforms all but one  competitive models in both IEMOCAP and EMO-DB in terms of accuracy.  In addition, our proposed model has the smallest size compared to  the previous competitive architectures.
 
 The rest of this paper is organized as follows: In Section 2 we mention some related works.  In Section 3,  we give a brief overview of quaternion operations and describe the proposed model. In Section 4  we detail our experiments and results. Finally, we draw the conclusion and give the future directions in Section 5.
 
\section{Relation to prior works}
A majority of previous literature on SER methods consist of variations CNNs such as a 1d-CNN for waveform features, CNN-LSTMs, transfer CNN models \cite{issa2020speech}. In addition, many methods have also focused on RNNs \cite{lim2016speech, satt2017efficient} to discover new features. 
	In the context of quaternion models, \cite{isokawa2003quaternion} introduced the ability to generate quaternion-based dense layers. Extending this work, QCNNs were developed for color image processing \cite{zhu2018quaternion}. The significance of quaternion models in the context of signal processing was originally shown by  \cite{parcollet2018quaternion} for automatic speech recognition; however, this model encoded views of a time-frame frequency as quaternions.
	Our work extends previous literature in that we exploit the decibel visualization of Mel-spectrograms as RGB arrays to encode in the quaternion domain rather than encoding the waveform features as pure quaternions or using a real-valued model.

 \begin{figure}[t]
  \centering
  \includegraphics[width=.9\linewidth]{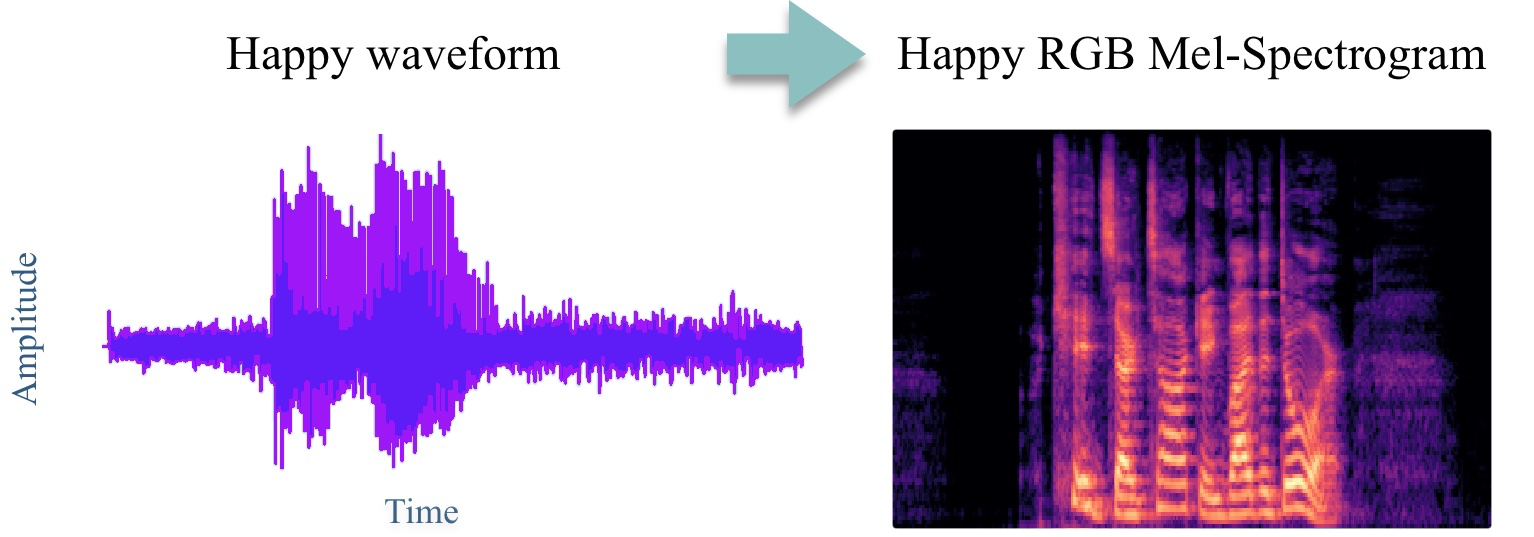}
  \caption{ The happy waveform converted to an RGB Mel-Spectrogram}
  \label{fig:slutask}
\end{figure}

\section{Model}

\subsection{Preprocessing}
Our feature generation pipeline consists of two parts: Mel-spectrogram array generation and RGB conversion.  
After the wav files were split into emotion-labelled folders, the Fourier transform  was computed on each of the speech waveforms to transform from the time domain to the frequency domain. While the time axis scaled linearly by seconds, the frequency axis was non-linearly transformed using a Mel transform, which is defined as:
\begin{equation}	
F_{\text{mel}}   = 2,595 \log_{10} (1+\frac{f}{700}) [dB]
\end{equation}
where $f$ denotes the frequency axis. The spectral amplitudes of the arrays were normalized between their dataset specific Max[dB] and Min [dB]. The mel-spectrogram was then converted to an image, with a color scale to represent the amplitudes (see example given in Figure 1). The RGB images were then split into  training ( 80\% ) and test  (20\%) datasets.

\subsection{Quaternion Convolutional Neural Networks}
\subsubsection {Quaternion Algebra}
Briefly, a quaternion is hypercomplex number which is an extension to complex numbers in 3-dimensional space with additionally imaginary parts $\pmb{j}$ and $\pmb{k}$. Thus, a quaternion $Q$ in domain $Q$ can be represented as $Q = r_0 + r_1\pmb{i} + r_2\pmb{j}+r_3\pmb{k}$ where $r \in R$ and $\pmb{i}, \pmb{j}$,  and $\pmb{k}$ are imaginary parts. Like complex number, Quaternions obey specific rules:
\begin{equation}
i^2=j^2=k^2=ijk=-1
\end{equation}
and the adddition is defined as
\begin{equation}
\begin{split}
Qa+Qb &= (r_{a_0} + r_{b_0})
\\&+ (r_{a_1} + r_{b1})\pmb{i} \\&+ (r_{a_2} + r_{b_2})\pmb{j} \\&+ (r_{a_3} + r_{b_3})\pmb{k}.
\end{split}
\end{equation}
Scalar multiplication is defined as
\begin{equation}
xQ = xr_0 + xr_1\pmb{i} + xr_2\pmb{j}+xr_3\pmb{k}.
\end{equation}
 \begin{figure}[h]
  \centering
  \includegraphics[width=1\linewidth]{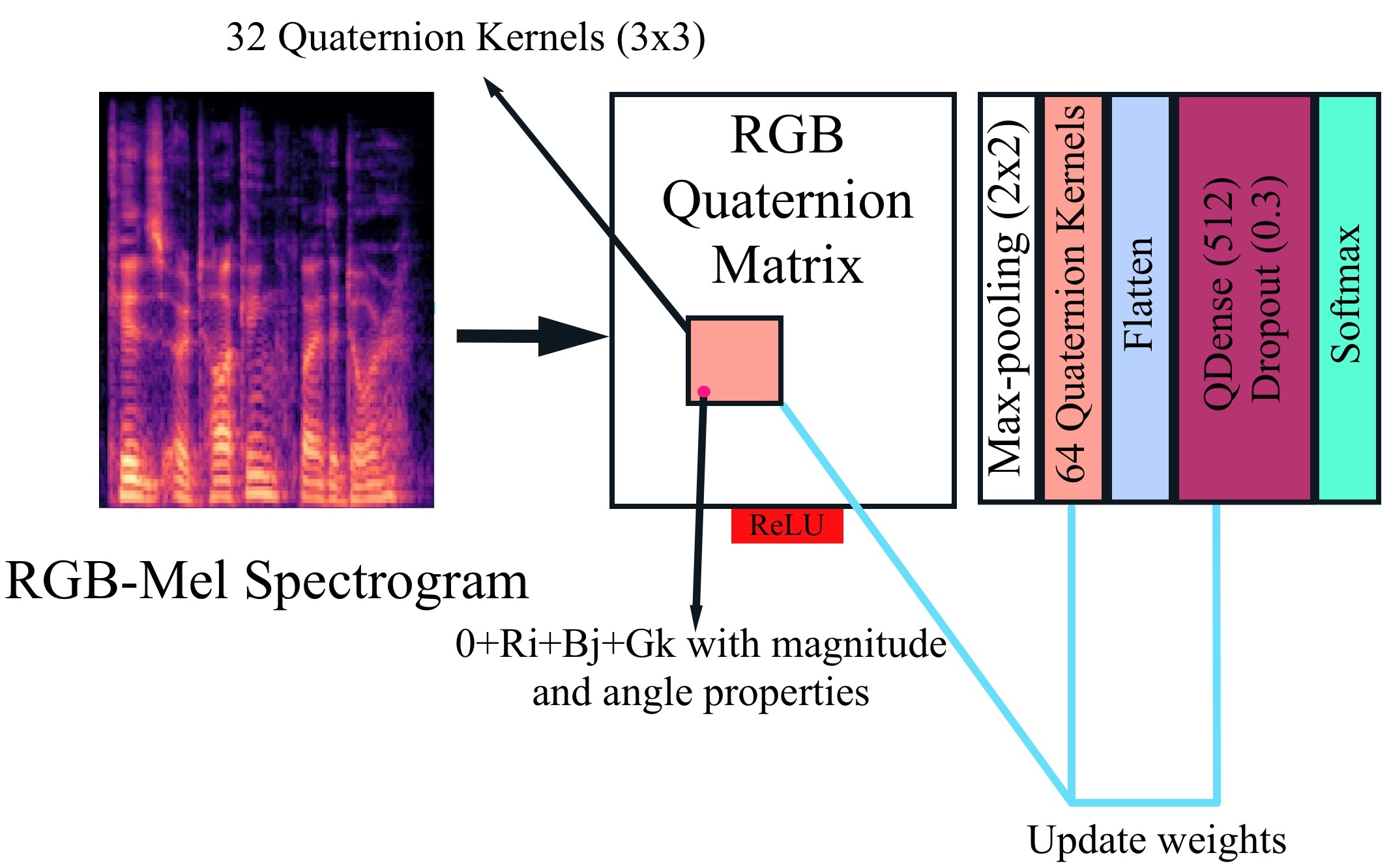}
  \caption{A  high-level block diagram of the proposed QCNN SER  system.}
  \label{fig:accuracyvsepoch}
\end{figure}
Element muultiplication is defined as:
\begin{equation}
\begin{split}
 Q_a \times Q_b & =( r_{a_0} r_{b_0} - r_{a_1} r_{b_1} - r_{a_2} r_{b_2}  - r_{a_3} r_{b_3}) \\&+ (r_{a_0} r_{b_1} + r_{a_1} r_{b_0}+ r_{a_2} r_{b_3} - r_{a_3} r_{b_2})\pmb{i} \\&+ (r_{a_0} r_{b_2} -  r_{a_1} r_{b_1}+ r_{a_2} r_{b_0}+ r_{a_3} r_{b_1})\pmb{j} 
 \\&+ (r_{a_0} r_{b_3} + r_{a_1} r_{b_2}- r_{a_2} r_{b_1}+ r_{a_3} r_{b_0}) \pmb{k},
\end{split}
\end{equation}
and rotation along axis $p$ is defined as 
\begin{equation}
\hat{R}=p\hat{Q}p^{-1}
\end{equation}
where 
\begin{equation}
p = \cos [ \frac{\theta}{2} ] + (p_1 \times i + p_2 \times j+p_3 \times k) \sin  \frac{ \theta}{2},   
\end{equation}
\begin{equation}
  \hat{Q}  =r_{q_0} + r_{q_1} \times i+ r_{q_2} \times j+ r_{q_3} \times k,
\end{equation}

\subsubsection {Quaternion Convolution}
We implement a quaternion color kernel similar to the one  proposed in \cite{zhu2018quaternion}. We do this by defining the $50\times75$ colored Mel-spectrogram image in a 3D vector color space as a quaternion matrix  $\hat{C} = [\hat{c}_{nn'}] \in H^{50\times75} $. 
We can thus represent the color channels in $C$ as:
\begin{equation} 
\hat{C} = 0+Ri+Gj+Bk
\end{equation} 
where $R,G,$ and $ B \in R^{50*75}$ are the red, green, and blue channels, respectively. 
By representing a pixel in 3D vector color space, the proposed model also deploys a quaternion kernel convolving around the input, in which the element $\hat{w}_{ll'}$ of the kernel $\hat{W}$ can be defined as:

\begin{equation}
\hat{w}_{ll'}  =s_{ll'}  \Bigl(  \cos \frac{\theta_{ll'}}{2}  +(  \frac{\sqrt{3}}{3} (i+j+k)) \sin \frac{\theta_{ll'}}{2}\Bigr) 
\end{equation}
where $\theta_{u' } \in [-\pi, \pi]$ and  $s_{u'} \in R$. Thus the quaternion convolution operation $*$ can be expressed as:
\begin{equation}
\hat{C } \times \hat{W} =\hat{F} =  [ \hat{f}_{kk'}  ] \in H^{((50-L+1)\times(75-L+1) )}
\end{equation}
where $\hat{W}$ is a $L\times L$ quaternion kernel and $[\hat{f}_{kk' } ]$ is defined as: 
\begin{equation}
\hat{f}_{kk' }  =  \sum_{l=1} \sum_{l'=1} \frac{1}{s_{ll'}}  \hat{w}_{ll'} \hat{c}_{(k+l)(k+l')} \hat{w}_{ll'}. 
\end{equation}
By implementing quaternion convolution kernels to perform rotation and scaling operations, the model  finds better representative features in the decibel color waveform space than real-valued convolution kernels. 
Additionally, the real-valued kernels apply scaling and pixel transformation separately to the three axes of color and create single-channel feature maps; whereas the quaternion kernels can more intuitively capture the color space as a whole without any type of color information loss since the channels are interrelated as a 3D vector.

\subsubsection{Connecting Layers}

The quaternion convolution layer is connected to other typical layers to construct the fully connected neural network. Specifically, max-pooling is performed on the imaginary parts and ReLU is used to reset invalid quaternion vector rotations to the nearest point in color space. In addition, the imaginary parts of the output of a quaternion layer are represented as 3 real numbers, such that a real-valued and vectorized output can be obtained to be connected to real-valued fully-connected layer. A softmax layer is then connected to this real-valued layer to train the QCNN model.

\subsubsection{Model Architecture}
Figure 2 shows a block diagram of the proposed model. The RGB-Mel spectrogram is transformed into a 3D color vector space with rotation magnitudes for the initial input layer. This layer is then followed by the quaternion convolutional layer, which consists of 32 quaternion kernels (all quaternion kernels, proposed in this paper, cover a 3x3 2D coordinate space) that perform rotation and scaling operations. ReLU is used to reset invalid quaternion vector rotations to the nearest point in color space. Afterward, max-pooling is performed on the imaginary color properties. This is followed by 64 quaternion kernels. The layer is then flattened to vectorize the quaternion layer to a real-valued layer. The real-valued dense layer is used to train the model by computing the gradient of the categorical cross-entropy loss function (which operates on softmax outputs for the classes of emotion).

\subsection{Training}

Weights are initialized as an imaginary quaternion with a uniform distribution in the interval [0, 1]. The imaginary unit is then normalized, with well-known criteria described by \cite{glorot2010understanding} to generate the quaternion weight $\hat{W}$ which was already defined above. Backpropagation is used to update the quaternion weights by applying the quaternion chain rule. Additionally, the loss function can be computed on the real-valued layer; specifically, the categorical cross-entropy function is used. Furthermore, the model used Adam as its stochastic optimizer and overfitting was addressed with the use of dropout (with a probability of 0.3). The model was trained for 50 epochs with 9 steps per validation epoch. The full model was implemented in Python with Tensorflow 2.3.0 as the deep learning backend.

  \section{Experiments}
  In order to measure the accuracy and effectivity of the model, QCNN is tested and compared against the most competitive models in SER on the RAVDESS, IEMOCAP, and EMODB datasets. In addition, we also compared the size of the models. 
  
 \subsection{Speech Datasets}
\subsubsection{RAVDESS}
The Ryerson Audio-Visual Database of Emotional Speech and Song (RAVDESS) \cite{livingstone2018ryerson} consists of speech samples of 8 categorical emotions— calm, happy, sad, angry, fearful, surprise, and disgust —spoken by 24 professional actors, vocalizing lexically-matched statements in a neutral North American accent. 
The classification accuracy of QCNN for the 8 classes of emotion on RAVDESS was compared with results from CNN-LSTM \cite{parry2019analysis}, a pre-feature extracted SVM \cite{shegokar2016continuous}, GResNets \cite{zeng2019spectrogram}, a controlled human accuracy study [RAVDESS], and Deep-CNN \cite{issa2020speech}. The results are summarized in table I.

\subsubsection{IEMOCAP}
The Interactive Emotional Dyadic Motion Capture (IEMOCAP) \cite{busso2008iemocap} database is an acted, multimodal and multispeaker database. It consists of 12 hours of audiovisual data, including video, speech, motion capture of face, text transcriptions. This paper makes use of the speech samples of 4 categorical emotions— anger, happiness, sadness, and neutrality —spoken by 10 (5 female and 5 male) professional actors in fluent English. 
The classification accuracy of QCNN for the 4 classes of emotion on IEMOCAP was compared with results from CNN-LSTM \cite{parry2019analysis}, DialogueRNN \cite{majumder2019dialoguernn}, BiF-AGRU \cite{li2019combining}, Deep-CNN \cite{issa2020speech}, DialogueGCN \cite{ghosal2019dialoguegcn}, BiERU-lc \cite{li2019improved}, and DSCNN \cite{kwon2020cnn}. The results are summarized in table II. 

\subsubsection{EMO-DB}
The Berlin EMO-DB \cite{burkhardt2005database} consists of speech samples representing 7 categorical emotions —anger, happiness, sadness, fear, disgust, boredom, and neutral speech—spoken by 10 professional actors (5 female and 5 male) in German. Each speaker acted 7 emotions for 10 different utterances (5 short and 5 long) with emotionally neutral linguistic contents. 
The classification accuracy for the 7 classes of emotion on EMO-DB was compared for the Multi-task LSTM \cite{goel2020cross}, RNN \cite{parry2019analysis}, CNN-LSTM \cite{parry2019analysis}, and Deep-CNN \cite{issa2020speech} methods with QCNN. The results are summarized in table III.

\subsection{Model Size}
The only model sizes that were reported for above models are Deep-CNN and CNN-LSTM. However, both being competitive models, they can be used as a bench mark to compare the QCNN model size. The results are summarized in Table IV.

\begin{table}[t]
  \caption{Classifier performance on RAVDESS}
   \label{tab:resultsI}
 \centering
\rowcolors{3}{green!25}{yellow!50}
\begin{tabular}{ *5l }    \toprule
\emph{Model} & \emph{Accuracy (\%)}   \\\midrule
CNN-LSTM \cite{parry2019analysis}  & 53.08 \\ 
Feature SVM& 60.10 \\ 
GResNets \cite{zeng2019spectrogram}   & 65.97 \\
Human Accuracy \cite{livingstone2018ryerson}  & 67.00  \\
Deep-CNN \cite{issa2020speech}  & 71.67  \\
QCNN [ours] & 77.87\\\bottomrule
 \hline
\end{tabular}
\end{table}

\subsection{Discussion}
The QCNN model presented yields state-of-the-art results on RAVDESS by \~6\%. Although the model does not achieve state-of-the-art results on EMO-DB or IEMOCAP, in both cases it underperforms the most competitive model by only $<$\~2\%.
	The balance of accuracy over model size is also considered in this paper, and although the model slightly underperforms in the IEMPCAP and EMO-DB datasets, the size of the model is significantly less. This is attributed to the ability to store RGB arrays in a quaternion domain to reduce model size [12]. However, in this paper, we did not explore encoding other features of waveforms such as Mel-frequency cepstral coefficients (MFCC), chromograms, spectral contrast features, or Tonnetz representations as pure quaternions. It is possible that separate Quaternion models encoding these features, separately or individually, could outperform the model presented in this paper. 
	Additionally, it is possible that an auxiliary neural network can be used in conjunction with the QCNN to capture higher-level features. General performance can also be improved by using effective waveform data augmentation techniques. 
	This research can also be extended to real-time SER for quaternion blocks to compute and process on speech-waveforms. However, for such a model it is expected to take advantage of other features rather than an RGB domain of Mel-spectrograms because of delta-computational expense.

\begin{table}[t]
  \caption{Classifier performance on IEMOCAP}
   \label{tab:resultsII}
 \centering
\rowcolors{3}{green!25}{yellow!50}
\begin{tabular}{ *5l }    \toprule
\emph{Model} & \emph{ Unweighted Accuracy (\%)}   \\\midrule
CNN-LSTM \cite{parry2019analysis}  & 50.17 \\ 
DialogueRNN \cite{majumder2019dialoguernn}& 63.40 \\ 
BiF-AGRU \cite{li2019combining}   & 63.50 \\
Deep-CNN \cite{issa2020speech}  & 64.30  \\
DialogueGCN \cite{ghosal2019dialoguegcn}  & 65.25  \\
BiERU-1c \cite{li2019improved}  & 66.11  \\
QCNN [ours]   & 70.46  \\
DSCNN DSCNN\cite{kwon2020cnn}& 72.00\\\bottomrule
 \hline
\end{tabular}
\end{table}

\section{Conclusion}
Current SER research is a dynamic, complex task relying on feature extraction and computational classification. To capture this highly complex qualitative information from speech waveforms, multiple feature extraction and classification techniques have been introduced in literature. However, many machine learning-based methods focus on higher-level features in the real-valued space. This paper reports a unique approach to feature and network encoding by using a quaternion structural model. Specifically, we encode the RGB domain of Mel-spectrogram features in a quaternion input and use custom quaternion convolutional layers to learn features in a quaternion space. 
These layers are implemented in a standard neural network structure to train on benchmark datasets such as RAVDESS, IEMOCAP, and EMO-DB. QCNN is reported to yield an accuracy of 77.87\%, 70.46\%, and 88.78\% for the RAVDESS, IEMOCAP, and EMO-DB datasets, respectfully. 
In comparison to other competitive methods, QCNN achieves state-of-the-art results on RAVDESS, and underperforms only one method in both IEMOCAP and EMO-DB. Additionally, QCNN is able to exploit its quaternion encoding for a reduced model size, confirming previous literature on the topic as well as providing an opportunity for deployment on lightweight machines such as voice-assistant devices. 
	Given the significant results in both accuracy and computational complexity, QCNN is expected to be a foundational block for the continuation of SER research.

\begin{table}[t]
  \caption{Classifier performance on EMO-DB}
   \label{tab:resultsIII}
 \centering
\rowcolors{3}{green!25}{yellow!50}
\begin{tabular}{ *5l }    \toprule
\emph{Model} & \emph{ Unweighted Accuracy (\%)}   \\\midrule
Multi-task LSTM \cite{goel2020cross}  & 58.14 \\ 
RNN\cite{parry2019analysis}& 63.21 \\ 
CNN-LSTM \cite{parry2019analysis}   & 69.72\\
QCNN [ours]   & 88.78  \\
Deep-CNN \cite{issa2020speech}& 90.01\\\bottomrule
 \hline
\end{tabular}
\end{table}

\begin{table}[t]
  \caption{Model sizes (Mb)}
   \label{tab:resultsIV}
 \centering
\rowcolors{3}{green!25}{yellow!50}
\begin{tabular}{ *5l }    \toprule
\emph{Model} & RAVDESS &EMO-DB&IEMOCAP   \\\midrule
Deep CNN & 74.5&69.5&88.5 \\ 
CNN-LSTM& 111.1 &94.4&128.3\\ 
QCNN& 42&31.2&67.7\\\bottomrule
 \hline
\end{tabular}
\end{table}

\bibliographystyle{IEEEtran}

\bibliography{mybib}

\begin{thebibliography}{10}
\providecommand{\url}[1]{#1}
\csname url@samestyle\endcsname
\providecommand{\newblock}{\relax}
\providecommand{\bibinfo}[2]{#2}
\providecommand{\BIBentrySTDinterwordspacing}{\spaceskip=0pt\relax}
\providecommand{\BIBentryALTinterwordstretchfactor}{4}
\providecommand{\BIBentryALTinterwordspacing}{\spaceskip=\fontdimen2\font plus
\BIBentryALTinterwordstretchfactor\fontdimen3\font minus
  \fontdimen4\font\relax}
\providecommand{\BIBforeignlanguage}[2]{{%
\expandafter\ifx\csname l@#1\endcsname\relax
\typeout{** WARNING: IEEEtran.bst: No hyphenation pattern has been}%
\typeout{** loaded for the language `#1'. Using the pattern for}%
\typeout{** the default language instead.}%
\else
\language=\csname l@#1\endcsname
\fi
#2}}
\providecommand{\BIBdecl}{\relax}
\BIBdecl

\bibitem{rethage2018wavenet}
D.~Rethage, J.~Pons, and X.~Serra, ``A wavenet for speech denoising,'' in
  \emph{2018 IEEE International Conference on Acoustics, Speech and Signal
  Processing (ICASSP)}.\hskip 1em plus 0.5em minus 0.4em\relax IEEE, 2018, pp.
  5069--5073.

\bibitem{benzeghiba2007automatic}
M.~Benzeghiba, R.~De~Mori, O.~Deroo, S.~Dupont, T.~Erbes, D.~Jouvet,
  L.~Fissore, P.~Laface, A.~Mertins, C.~Ris \emph{et~al.}, ``Automatic speech
  recognition and speech variability: A review,'' \emph{Speech communication},
  vol.~49, no. 10-11, pp. 763--786, 2007.

\bibitem{ramakrishnan2013speech}
S.~Ramakrishnan and I.~M. El~Emary, ``Speech emotion recognition approaches in
  human computer interaction,'' \emph{Telecommunication Systems}, vol.~52,
  no.~3, pp. 1467--1478, 2013.

\bibitem{ashar2020speaker}
A.~Ashar, M.~S. Bhatti, and U.~Mushtaq, ``Speaker identification using a hybrid
  cnn-mfcc approach,'' in \emph{2020 International Conference on Emerging
  Trends in Smart Technologies (ICETST)}.\hskip 1em plus 0.5em minus
  0.4em\relax IEEE, 2020, pp. 1--4.

\bibitem{an2017emotional}
S.~An, Z.~Ling, and L.~Dai, ``Emotional statistical parametric speech synthesis
  using lstm-rnns,'' in \emph{2017 Asia-Pacific Signal and Information
  Processing Association Annual Summit and Conference (APSIPA ASC)}.\hskip 1em
  plus 0.5em minus 0.4em\relax IEEE, 2017, pp. 1613--1616.

\bibitem{pascual2017segan}
S.~Pascual, A.~Bonafonte, and J.~Serra, ``Segan: Speech enhancement generative
  adversarial network,'' \emph{arXiv preprint arXiv:1703.09452}, 2017.

\bibitem{humphrey2012learning}
E.~J. Humphrey, T.~Cho, and J.~P. Bello, ``Learning a robust tonnetz-space
  transform for automatic chord recognition,'' in \emph{2012 IEEE International
  Conference on Acoustics, Speech and Signal Processing (ICASSP)}.\hskip 1em
  plus 0.5em minus 0.4em\relax IEEE, 2012, pp. 453--456.

\bibitem{palaz2015convolutional}
D.~Palaz, M.~M. Doss, and R.~Collobert, ``Convolutional neural networks-based
  continuous speech recognition using raw speech signal,'' in \emph{2015 IEEE
  International Conference on Acoustics, Speech and Signal Processing
  (ICASSP)}.\hskip 1em plus 0.5em minus 0.4em\relax IEEE, 2015, pp. 4295--4299.

\bibitem{miao2015eesen}
Y.~Miao, M.~Gowayyed, and F.~Metze, ``Eesen: End-to-end speech recognition
  using deep rnn models and wfst-based decoding,'' in \emph{2015 IEEE Workshop
  on Automatic Speech Recognition and Understanding (ASRU)}.\hskip 1em plus
  0.5em minus 0.4em\relax IEEE, 2015, pp. 167--174.

\bibitem{graves2013hybrid}
A.~Graves, N.~Jaitly, and A.-r. Mohamed, ``Hybrid speech recognition with deep
  bidirectional lstm,'' in \emph{2013 IEEE workshop on automatic speech
  recognition and understanding}.\hskip 1em plus 0.5em minus 0.4em\relax IEEE,
  2013, pp. 273--278.

\bibitem{wollmer2010combining}
M.~W{"o}llmer, B.~Schuller, F.~Eyben, and G.~Rigoll, ``Combining long
  short-term memory and dynamic bayesian networks for incremental
  emotion-sensitive artificial listening,'' \emph{IEEE Journal of selected
  topics in signal processing}, vol.~4, no.~5, pp. 867--881, 2010.

\bibitem{isokawa2003quaternion}
T.~Isokawa, T.~Kusakabe, N.~Matsui, and F.~Peper, ``Quaternion neural network
  and its application,'' in \emph{International conference on knowledge-based
  and intelligent information and engineering systems}.\hskip 1em plus 0.5em
  minus 0.4em\relax Springer, 2003, pp. 318--324.

\bibitem{zhu2018quaternion}
X.~Zhu, Y.~Xu, H.~Xu, and C.~Chen, ``Quaternion convolutional neural
  networks,'' in \emph{Proceedings of the European Conference on Computer
  Vision (ECCV)}, 2018, pp. 631--647.

\bibitem{issa2020speech}
D.~Issa, M.~F. Demirci, and A.~Yazici, ``Speech emotion recognition with deep
  convolutional neural networks,'' \emph{Biomedical Signal Processing and
  Control}, vol.~59, p. 101894, 2020.

\bibitem{lim2016speech}
W.~Lim, D.~Jang, and T.~Lee, ``Speech emotion recognition using convolutional
  and recurrent neural networks,'' in \emph{2016 Asia-Pacific signal and
  information processing association annual summit and conference
  (APSIPA)}.\hskip 1em plus 0.5em minus 0.4em\relax IEEE, 2016, pp. 1--4.

\bibitem{satt2017efficient}
A.~Satt, S.~Rozenberg, and R.~Hoory, ``Efficient emotion recognition from
  speech using deep learning on spectrograms.'' in \emph{Interspeech}, 2017,
  pp. 1089--1093.

\bibitem{parcollet2018quaternion}
T.~Parcollet, Y.~Zhang, M.~Morchid, C.~Trabelsi, G.~Linar{`e}s, R.~De~Mori, and
  Y.~Bengio, ``Quaternion convolutional neural networks for end-to-end
  automatic speech recognition,'' \emph{arXiv preprint arXiv:1806.07789}, 2018.

\bibitem{brigham1967fast}
E.~O. Brigham and R.~Morrow, ``The fast fourier transform,'' \emph{IEEE
  spectrum}, vol.~4, no.~12, pp. 63--70, 1967.

\bibitem{glorot2010understanding}
X.~Glorot and Y.~Bengio, ``Understanding the difficulty of training deep
  feedforward neural networks,'' in \emph{Proceedings of the thirteenth
  international conference on artificial intelligence and statistics}, 2010,
  pp. 249--256.

\bibitem{livingstone2018ryerson}
S.~R. Livingstone and F.~A. Russo, ``The ryerson audio-visual database of
  emotional speech and song (ravdess): A dynamic, multimodal set of facial and
  vocal expressions in north american english,'' \emph{PloS one}, vol.~13,
  no.~5, p. e0196391, 2018.

\bibitem{parry2019analysis}
J.~Parry, D.~Palaz, G.~Clarke, P.~Lecomte, R.~Mead, M.~Berger, and G.~Hofer,
  ``Analysis of deep learning architectures for cross-corpus speech emotion
  recognition.'' in \emph{INTERSPEECH}, 2019, pp. 1656--1660.

\bibitem{shegokar2016continuous}
P.~Shegokar and P.~Sircar, ``Continuous wavelet transform based speech emotion
  recognition,'' in \emph{2016 10th International Conference on Signal
  Processing and Communication Systems (ICSPCS)}.\hskip 1em plus 0.5em minus
  0.4em\relax IEEE, 2016, pp. 1--8.

\bibitem{zeng2019spectrogram}
Y.~Zeng, H.~Mao, D.~Peng, and Z.~Yi, ``Spectrogram based multi-task audio
  classification,'' \emph{Multimedia Tools and Applications}, vol.~78, no.~3,
  pp. 3705--3722, 2019.

\bibitem{busso2008iemocap}
C.~Busso, M.~Bulut, C.-C. Lee, A.~Kazemzadeh, E.~Mower, S.~Kim, J.~N. Chang,
  S.~Lee, and S.~S. Narayanan, ``Iemocap: Interactive emotional dyadic motion
  capture database,'' \emph{Language resources and evaluation}, vol.~42, no.~4,
  p. 335, 2008.

\bibitem{majumder2019dialoguernn}
N.~Majumder, S.~Poria, D.~Hazarika, R.~Mihalcea, A.~Gelbukh, and E.~Cambria,
  ``Dialoguernn: An attentive rnn for emotion detection in conversations,'' in
  \emph{Proceedings of the AAAI Conference on Artificial Intelligence},
  vol.~33, 2019, pp. 6818--6825.

\bibitem{li2019combining}
C.~Li, J.~Jiao, Y.~Zhao, and Z.~Zhao, ``Combining gated convolutional networks
  and self-attention mechanism for speech emotion recognition,'' in \emph{2019
  8th International Conference on Affective Computing and Intelligent
  Interaction Workshops and Demos (ACIIW)}.\hskip 1em plus 0.5em minus
  0.4em\relax IEEE, 2019, pp. 105--109.

\bibitem{ghosal2019dialoguegcn}
D.~Ghosal, N.~Majumder, S.~Poria, N.~Chhaya, and A.~Gelbukh, ``Dialoguegcn: A
  graph convolutional neural network for emotion recognition in conversation,''
  \emph{arXiv preprint arXiv:1908.11540}, 2019.

\bibitem{li2019improved}
Y.~Li, T.~Zhao, and T.~Kawahara, ``Improved end-to-end speech emotion
  recognition using self attention mechanism and multitask learning.'' in
  \emph{Interspeech}, 2019, pp. 2803--2807.

\bibitem{kwon2020cnn}
S.~Kwon \emph{et~al.}, ``A cnn-assisted enhanced audio signal processing for
  speech emotion recognition,'' \emph{Sensors}, vol.~20, no.~1, p. 183, 2020.

\bibitem{burkhardt2005database}
F.~Burkhardt, A.~Paeschke, M.~Rolfes, W.~F. Sendlmeier, and B.~Weiss, ``A
  database of german emotional speech,'' in \emph{Ninth European Conference on
  Speech Communication and Technology}, 2005.

\bibitem{goel2020cross}
S.~Goel and H.~Beigi, ``Cross lingual cross corpus speech emotion
  recognition,'' \emph{arXiv preprint arXiv:2003.07996}, 2020.

\end{thebibliography}


\end{document}